\documentclass[aps,prb,amsmath,amssymb,twocolumn,superscriptaddress,showpacs]{revtex4-1}
\usepackage{graphicx}% Include figure files
\usepackage{bm}% bold math
\usepackage{hyperref}
\usepackage{color}

\newcommand{\cbar}{\bar{c}}
\newcommand{\cD}{{\cal D}}
\newcommand{\cF}{{\cal F}}
\newcommand{\cFLP}{{\cal F}_{\rm LP}}
\newcommand{\cH}{{\cal H}} %OMIT
\newcommand{\cZ}{{\cal Z}}
\newcommand{\half}{\frac{1}{2}}
\newcommand{\kT}{k_{\rm B}T}
\newcommand{\Tc}{T_{\rm c}}
\newcommand{\trho}{\tilde{\rho}}
\newcommand{\tU}{\tilde{U}}
\newcommand{\tUmin}{\tilde{U}_{\rm min}}
\newcommand{\veck}{{\bf k}}
\newcommand{\vecr}{{\bf r}}

\begin{document}

\title{Stability of Quasicrystals Composed of Soft Isotropic Particles}

\author{Kobi Barkan}
\affiliation{Raymond and Beverly Sackler School of Physics and
  Astronomy, Tel Aviv University, 69978 Tel Aviv, Israel} 
\author{Haim Diamant}
\affiliation{Raymond and Beverly Sackler School of Chemistry, Tel Aviv
  University, 69978 Tel Aviv, Israel}  
\author{Ron Lifshitz}
\email[Corresponding author:\ ]{ronlif@tau.ac.il}
\affiliation{Raymond and Beverly Sackler School of Physics and
  Astronomy, Tel Aviv University, 69978 Tel Aviv, Israel} 

\date{March 15, 2011}

\begin{abstract} 
  Quasicrystals whose building blocks are of mesoscopic rather than
  atomic scale have recently been discovered in several soft-matter
  systems. Contrary to metallurgic quasicrystals whose source of
  stability remains a question of great debate to this day, we argue
  that the stability of certain soft-matter quasicrystals can be
  directly explained by examining a coarse-grained free energy for a
  system of soft isotropic particles. We show, both theoretically and
  numerically, that the stability can be attributed to the existence
  of two natural length scales in the pair potential, combined with
  effective three-body interactions arising from entropy.  Our newly
  gained understanding of the stability of soft quasicrystals allows
  us to point at their region of stability in the phase diagram, and
  thereby may help control the self-assembly of quasicrystals and a
  variety of other desired structures in future experimental
  realizations.
\end{abstract}


\maketitle

Quasicrystals are more common than one had originally expected when
their discovery was first announced.\cite{shechtman} More than a
hundred different metallic alloys are known to form stable
quasicrystalline phases of icosahedral symmetry alone,\cite{tsai08}
with a few dozen additional stable phases exhibiting decagonal
(10-fold) and possibly other symmetries.\cite{steurer04} Yet, to this
date, there is no general agreement regarding the origin of their
stability and the respective roles of energy and entropy in
determining the observed phases.\cite{deboissieu06,*henley06} These
growing numbers of stable solid-state quasicrystals, whose building
blocks are on the atomic scale, have been joined in recent years by a
host of soft-matter systems exhibiting quasiperiodic long-range order
with building blocks on a much larger scale of tens to hundreds of
nanometers---micelle-forming
dendrimers,\cite{zeng04,*zeng05,percec09a,*percec09b} star block
copolymers,\cite{takano05,*hayashida07} mesoporous
silica,\cite{[{C.~Xiao, K.~Miyasaka, N.~Fujita, Y.~Sakamoto, and
    O.~Terasaki, in {\it ICQ11---Book of Abstracts}, Y. Ishii and T.
    Ishimasa, eds. (Hokkaido University, Sapporo, 2010) p. 28.
    Reproduced with permission as figure 2 of }]Dubois11} and binary
systems of nanoparticles.\cite{talapin09} These newly discovered soft
quasicrystals hold the promise for applications based on
self-assembled nanomaterials,\cite{electronics,*selfassemblyreview}
with unique electronic or photonic properties that take advantage of
their quasiperiodicity.\cite{QCtoNano} At the same time, they provide
alternative experimental platforms for the basic study of
quasiperiodic long-range order, and offer the opportunity to study the
thermodynamic stability of quasicrystals from a fresh viewpoint.  To
this date, soft quasicrystals have been observed only with dodecagonal
point-group symmetry, having quasiperiodic order in the 12-fold plane
and periodic order normal to the plane, whereas dodecagonal
solid-state quasicrystals are rare and mostly only
\emph{metastable}.\cite{steurer04} Soft quasicrystals may belong,
therefore, to a distinct class of quasicrystals, whose source of
stability is likely to be different from their solid-state
counterparts. We propose here a simple theoretical framework to
address these new systems.  We use it to explain the stability of the
observed structures and indicate the (surprisingly simple) minimum
conditions under which quasicrystals could be stabilized.  Knowledge
of these conditions gives us the ability to estimate the location of
the region in the phase diagram where quasicrystals should be stable,
and thus may help control the self-assembly of quasicrystals and other
desired structures in future experimental realizations.

Several microscopic models have been studied over the years, mainly
using computer simulations, but also using sophisticated analytical
methods such as thermodynamic perturbation theory,\cite{Denton98} to
explore the structures arising from pair potentials that possess more
than one microscopic length scale. These studies have yielded
surprisingly rich phase diagrams even within the limited scope of
single-component systems, interacting via isotropic pair
potentials,\cite{Jagla99,Ziherl01,*Glaser07,Pauschenwein08,Saija09,%
  *Prestipino09,Shin09} in some cases even finding stable
quasicrystals.\cite{Denton98,Dzugutov93,Quandt99,*Skibinsky99,%
  Engel07,*Engel08,Keys07} On the other hand, phenomenological models
based on coarse-grained free energies have been widely applied to
treat phase diagrams and transitions\cite{alexander84} and to explain
the stability of different phases, including
quasicrystals.\cite{kkl,*Mermin85,*Gronlund88} This is especially true
in the case of soft-matter systems\cite{Gompper94,Dotera06,%
  *Dotera07} due to their intermediate mesoscopic building blocks,
which are significantly larger than the atomic scale, rendering a
long-wavelength gradient expansion a valid approximation.

A particular free energy of this sort, which is relevant for what
follows below, was developed by Lifshitz and Petrich,\cite[henceforth
LP]{faraday} who extended the Swift-Hohenberg
equation\cite{swift77,*Cross09} to study parametrically-excited
surface waves (Faraday waves),\cite{edwards93} exhibiting dodecagonal
quasiperiodic order. The LP free energy has the form
% Single column version for submission
% \begin{equation}
%  \label{eq:FLP}
%  \cFLP\left[\rho(\vecr)\right] = \int \!dx\, dy\, \left\{
%    \frac12 \left[(\nabla^2+1)(\nabla^2+q^2)\rho\right]^2
%   - \frac12 \varepsilon\rho^2
%    - \frac13 \alpha \rho^3 + \frac14 \rho^4 \right\},
% \end{equation}
%
%Two column version for preprint
\begin{eqnarray}
 \label{eq:FLP}\nonumber
 \cFLP\left[\rho(\vecr)\right] &=&\int \!dx\, dy\, \left\{
   \frac12 \left[(\nabla^2+1)(\nabla^2+q^2)\rho\right]^2\right.\\ 
  && \left.- \frac12 \varepsilon\rho^2
   - \frac13 \alpha \rho^3 + \frac14 \rho^4 \right\},
\end{eqnarray}
where $\nabla^2 = {\partial_x}^2 + {\partial_y}^2$ is the
2-dimensional Laplacian. It is quite generic, imposing only two
requirements on a material described by a 2-dimensional density
$\rho(x,y)$: (a) the existence of two characteristic length scales,
whose ratio is given by the parameter $q$; and (b) effective 3-body
interactions, weighted by the parameter $\alpha$, that act to
stabilize structures containing triplets of density modes with wave
vectors adding up to zero. LP showed that if $q$ is chosen around
$2\cos(\pi/12)=\sqrt{2+\sqrt3}\simeq 1.93$ one can obtain a
quasiperiodic ground state with dodecagonal symmetry, yet no choice of
$q$ yields globally-stable ground states with octagonal or decagonal
symmetry, due to insufficiently-many resonant triplets of modes.
Inspired by this simple result, we conjectured that the existence of
two characteristic length scales along with 3-body interactions may
constitute the source of stability of soft quasicrystals, all of which
(to date) are dodecagonal.\cite{LD07} Here we confirm this conjecture
by coarse-graining a microscopic partition function for isotropic soft
particles into an effective free energy. In the limit of small
deviations away from the uniform, or liquid, phase this coarse-grained
free energy can be expanded in a power series and mapped onto the
simple LP form~\eqref{eq:FLP}, allowing us to gain important insight
from the simpler LP model, and consequently to explain the stability
of the observed phases using the full coarse-grained free energy.

Our starting point is the grand partition function for a system of
particles with pairwise interactions,
\begin{subequations}
\begin{eqnarray}
  &&{\cZ} = \sum_{N=0}^\infty \frac{e^{\beta\mu N}}{N!} 
  \int\prod_{n=1}^N d\vecr_n e^{-\beta\cH[\{\vecr_n\}]},\\ 
  &&\cH[\{\vecr_n\}] = \half\sum_{m\neq n} U(\vecr_m-\vecr_n),
\end{eqnarray}
\end{subequations}
where $\{\vecr_n\}_{n=1,\ldots,N}$ are the 2-dimensional positions of
the centers of $N$ particles, $U$ their pair potential,
$\beta=(\kT)^{-1}$ the inverse temperature, and $\mu$ the chemical
potential, which determines the mean particle density, $\cbar$. Using
standard methods,\cite{Fredrickson06} one can rewrite the partition
function in terms of collective coordinates---namely, the particle
density, $c(\vecr)\equiv\sum_{n=1}^N \delta(\vecr-\vecr_n)$, and its
conjugate field---rather than discrete positions. At the mean-field
level, which amounts to a saddle-point approximation for the
integration over the conjugate field, the transformed partition
function becomes ${\cZ}=\int{\cD}c e^{-\beta\cF[c]}$, where
% Single column version for submission
% \begin{equation}
%   \cF\left[c(\vecr)\right] = \half\int d\vecr d\vecr'
%   c(\vecr)U(\vecr-\vecr')c(\vecr')  
%   +\int d\vecr \{\kT c(\vecr)[\ln c(\vecr)-1]-\mu c(\vecr)\}.
% \label{cF}
% \end{equation}
%
% Two column version for preprint
\begin{eqnarray}
  \cF\left[c(\vecr)\right] &=& \half\int d\vecr d\vecr' c(\vecr)U(\vecr-\vecr')c(\vecr')  
 \nonumber\\
  && +\int d\vecr \{\kT c(\vecr)[\ln c(\vecr)-1]-\mu c(\vecr)\}.
\label{cF}
\end{eqnarray}
The coarse-grained free energy functional given in Eq.~(\ref{cF})
contains the familiar mean-field terms of pair interaction and ideal
entropy. Although it could have been written from the outset, we wish
to highlight the ability to extend the current theory to higher order,
particularly in light of the cautionary remarks of Schwartz and
Vinograd.\cite{Schwartz02} We shall assume that the equilibrium
density field is the one that minimizes $\cF$ for the given $T$ and
$\mu$ and the specific choice of $U(\vecr)$. Results of such direct
minimization will be presented shortly. However, since it is not {\it
  a priori} obvious what pair potentials and thermodynamic parameters
may yield quasicrystalline order, it is beneficial first to
characterize $\cF$ and relate it to $\cFLP$ of Eq.\ (\ref{eq:FLP}).

Above a certain critical temperature, $T>\Tc$, the equilibrium density
for any $\mu$ should be uniform, $c(\vecr)\equiv\cbar$.  Minimizing
$\cF$ with respect to such a uniform field yields the relation, 
\begin{equation}
  \label{eq:cbar}
  \mu = \kT\ln\cbar + \tU_0\cbar, 
\end{equation}
where $\tU_0\equiv\int d\vecr U(\vecr)$. For $T<\Tc$ the equilibrium
density is expected to become nonuniform at a certain value of $\mu$
(or, alternatively, above a certain mean density $\cbar$). Assuming
that $T$ is only slightly smaller than $\Tc$, we substitute
$c(\vecr)=\cbar[1+\rho(\vecr)]$ in $\cF$ and expand to 4th order in
small $\rho$. The result can be written as
\begin{eqnarray}
  \label{eq:Fexpansion}\nonumber
  &&\frac{\cF\left[\rho\right]}{\cbar\kT_{\rm c}} \simeq
  \int\frac{d\veck}{(2\pi)^2} 
  \half \frac{\tU(\veck)-\tUmin}{|\tUmin|} \left|\trho(\veck)\right|^2\\
  &&+
  \int d\vecr \left\{ \half \frac{T-\Tc}{\Tc} \rho^2(\vecr) 
  - \frac{1}{6}\frac{T}{\Tc} \rho^3(\vecr)
  + \frac{1}{12}\frac{T}{\Tc} \rho^4(\vecr) \right\},\quad
\end{eqnarray}
where tildes denote Fourier-transformed quantities. The critical
temperature, below which perturbations start to grow, is given by
\begin{equation}
  \label{eq:TC}
\kT_{\rm c}=-\cbar\tUmin,
\end{equation}
where $\tUmin$ is the minimum of the Fourier-transformed pair
potential, which must be negative for $\Tc$ to be positive.  The
approximate coarse-grained free energy~(\ref{eq:Fexpansion}) resembles
$\cFLP$~(\ref{eq:FLP}), with the gradient term replaced by a
Fourier-space integral of the pair potential, and the expansion in
powers of the density arising from the entropy term of Eq.~(\ref{cF}).

%##################### Figure 1 #################################### 
\begin{figure*}[t]
\begin{center}
\scalebox{0.215}{\rotatebox{00}{\includegraphics{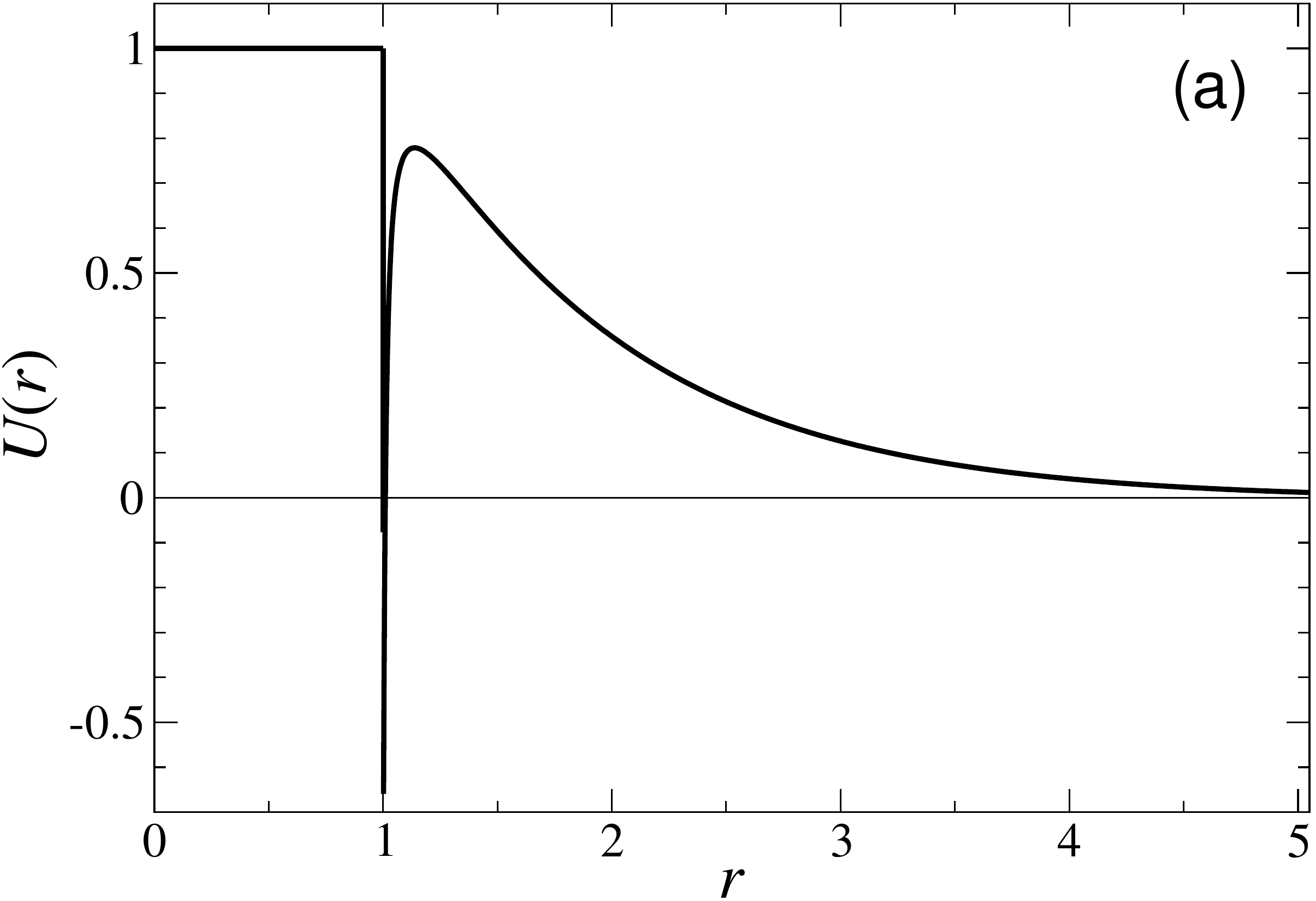}}}
\scalebox{0.215}{\rotatebox{00}{\includegraphics{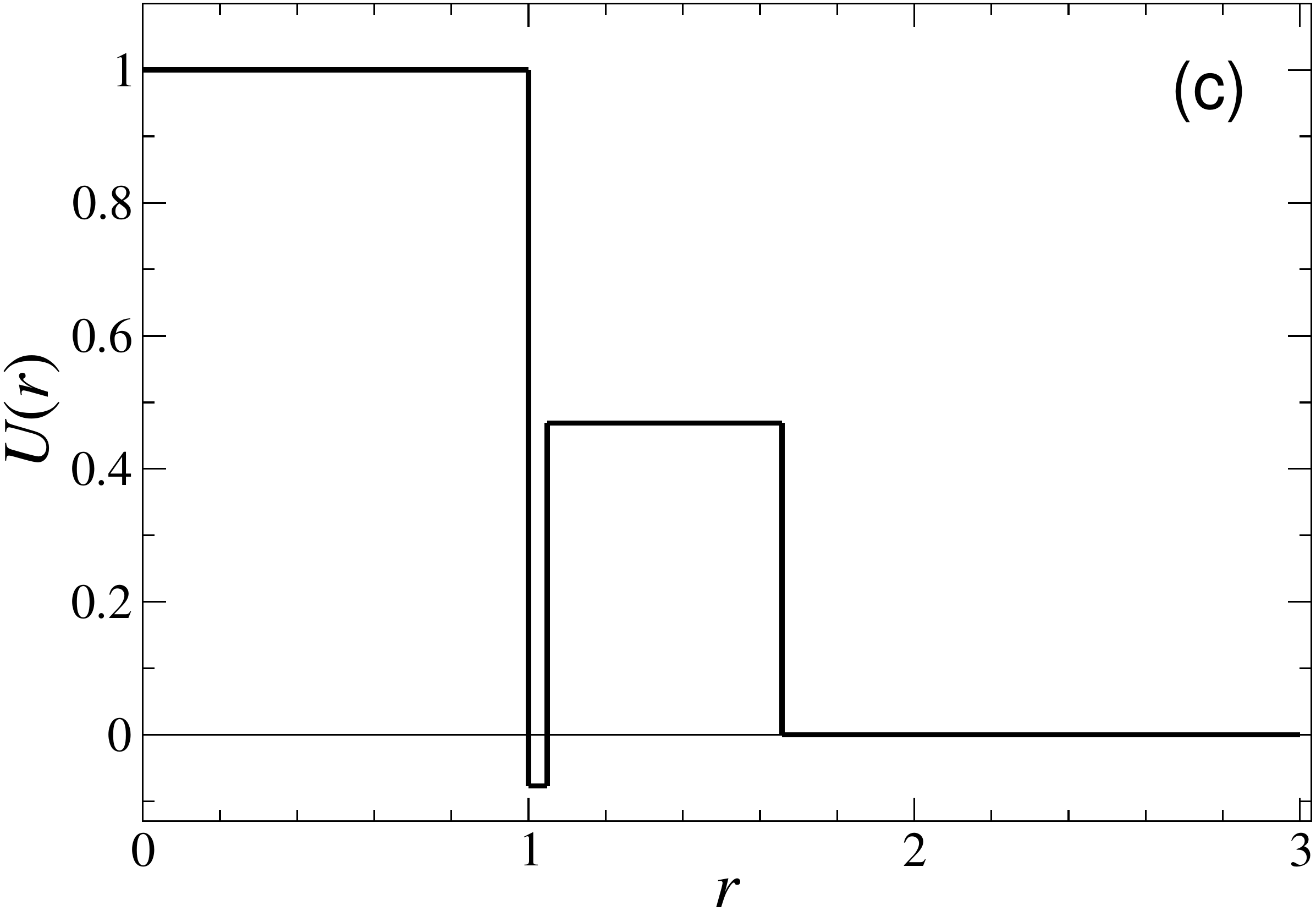}}}
\scalebox{0.215}{\rotatebox{00}{\includegraphics{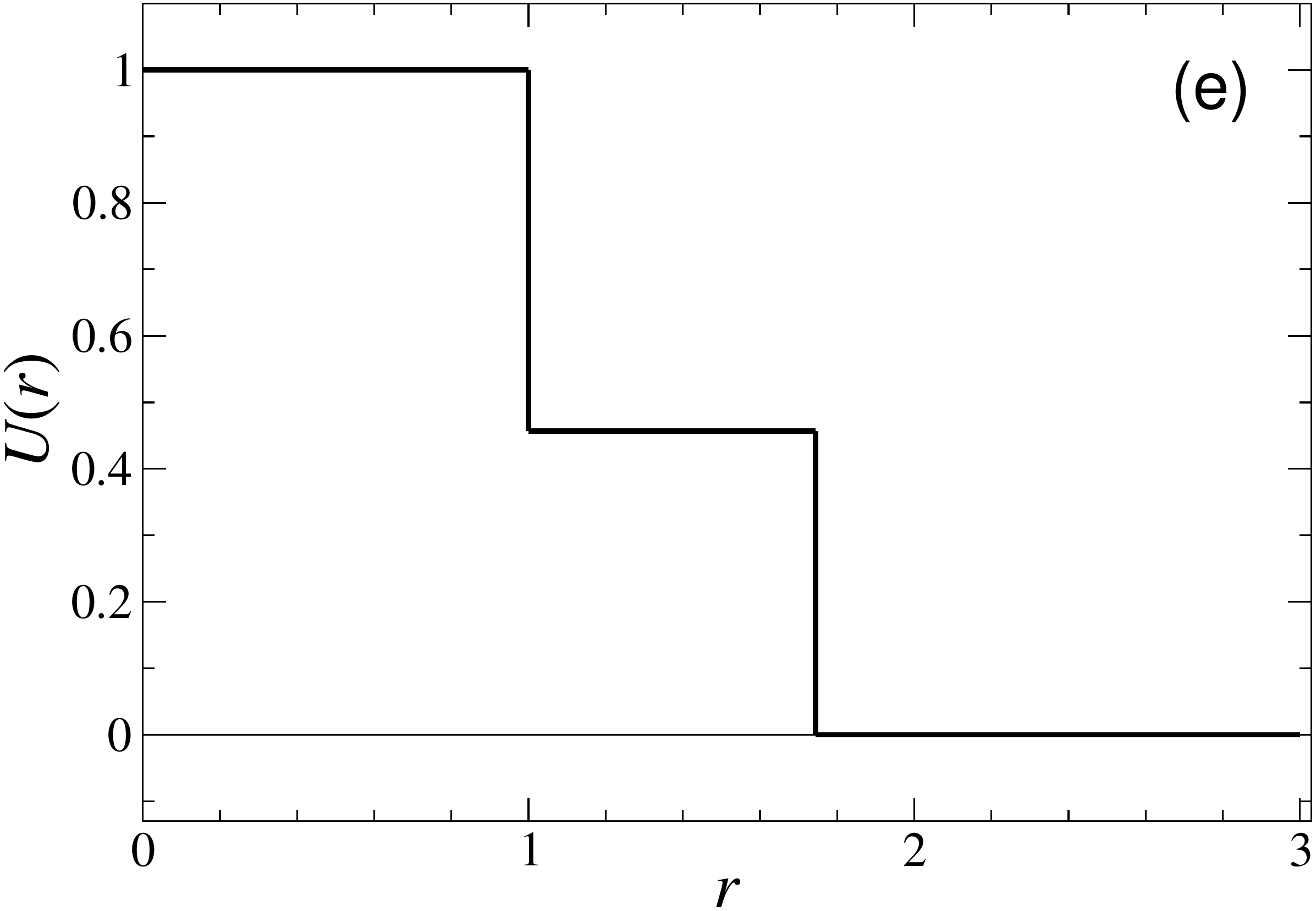}}}\\
\scalebox{0.215}{\rotatebox{00}{\includegraphics{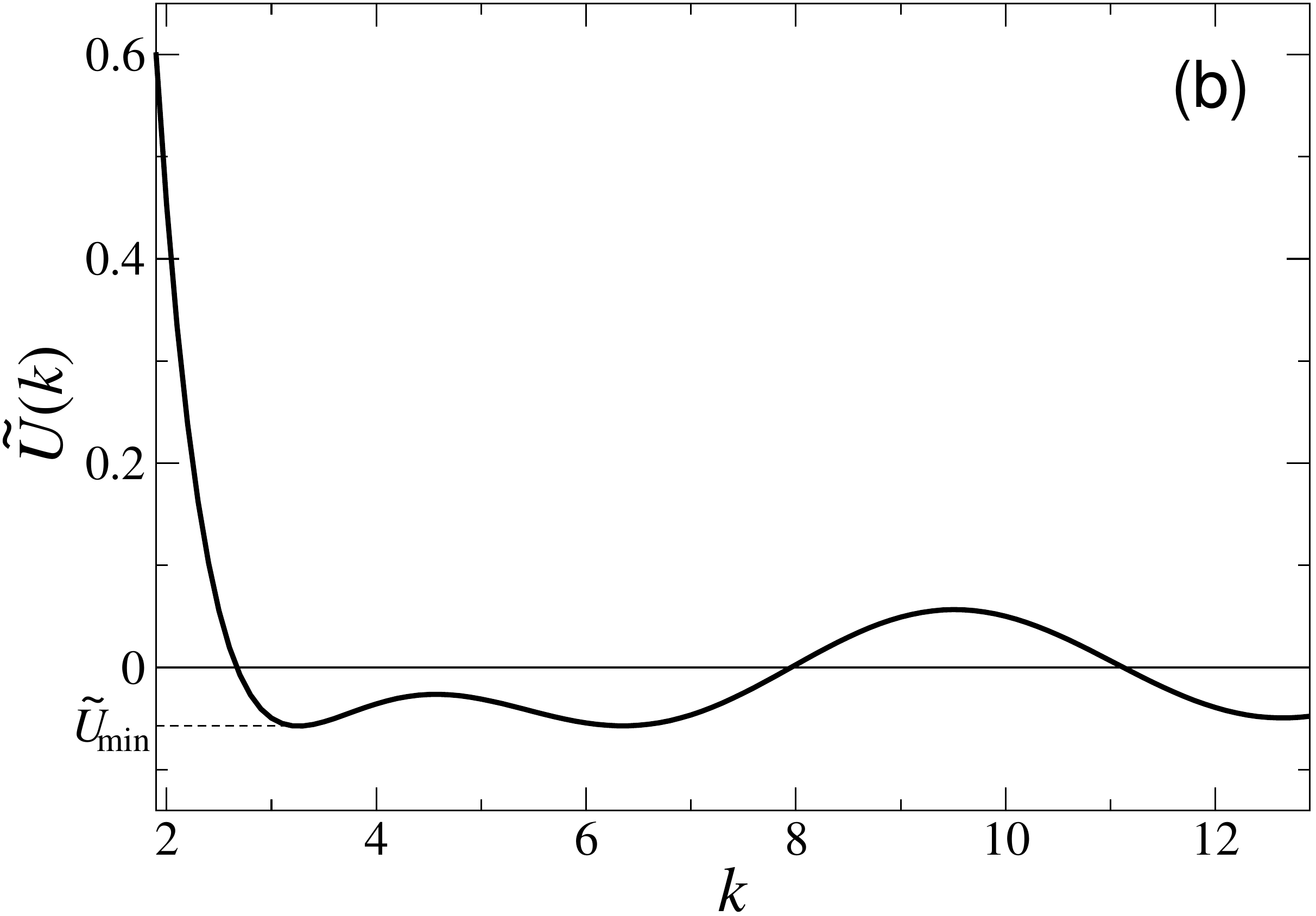}}}
\scalebox{0.215}{\rotatebox{00}{\includegraphics{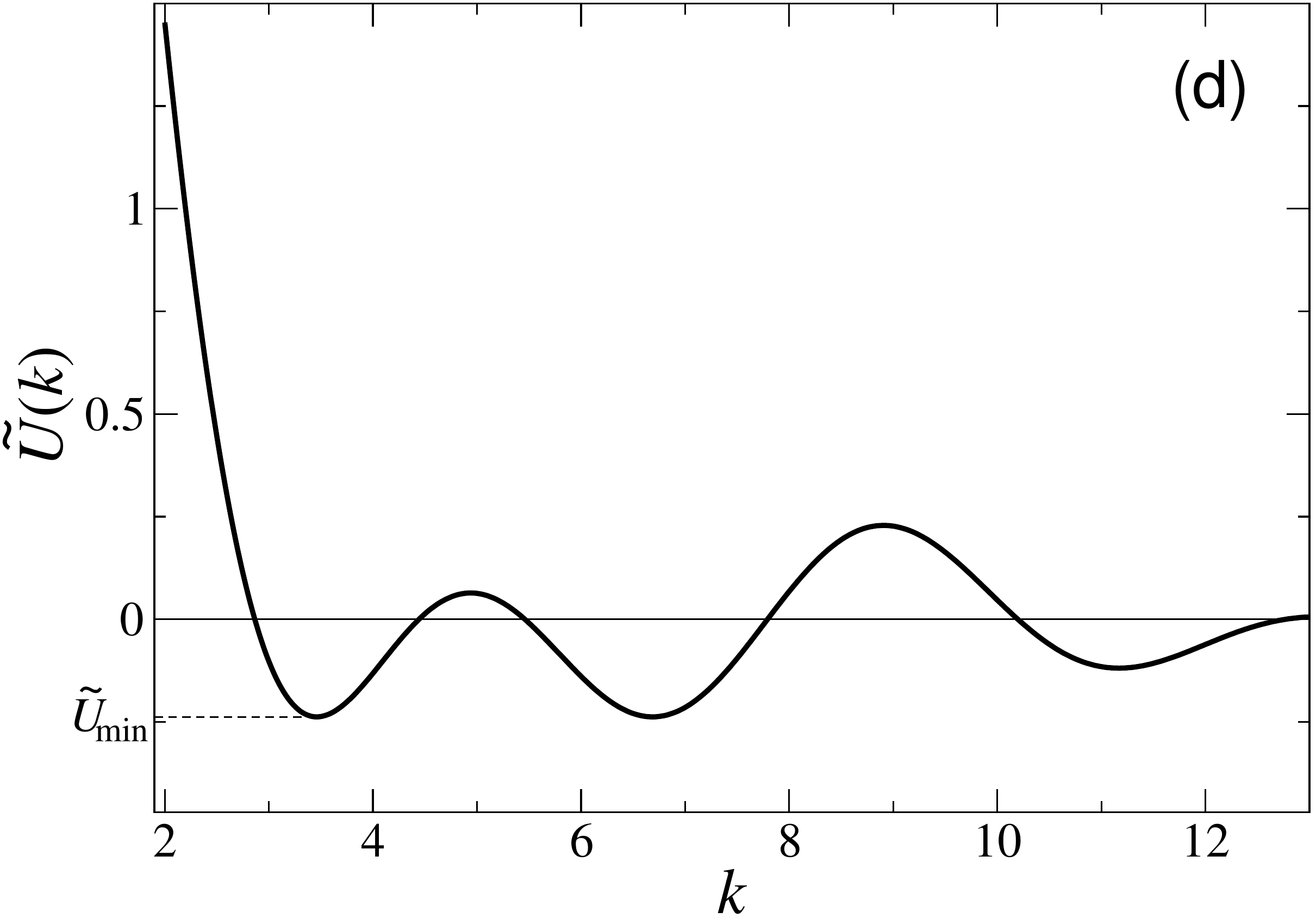}}}
\scalebox{0.215}{\rotatebox{00}{\includegraphics{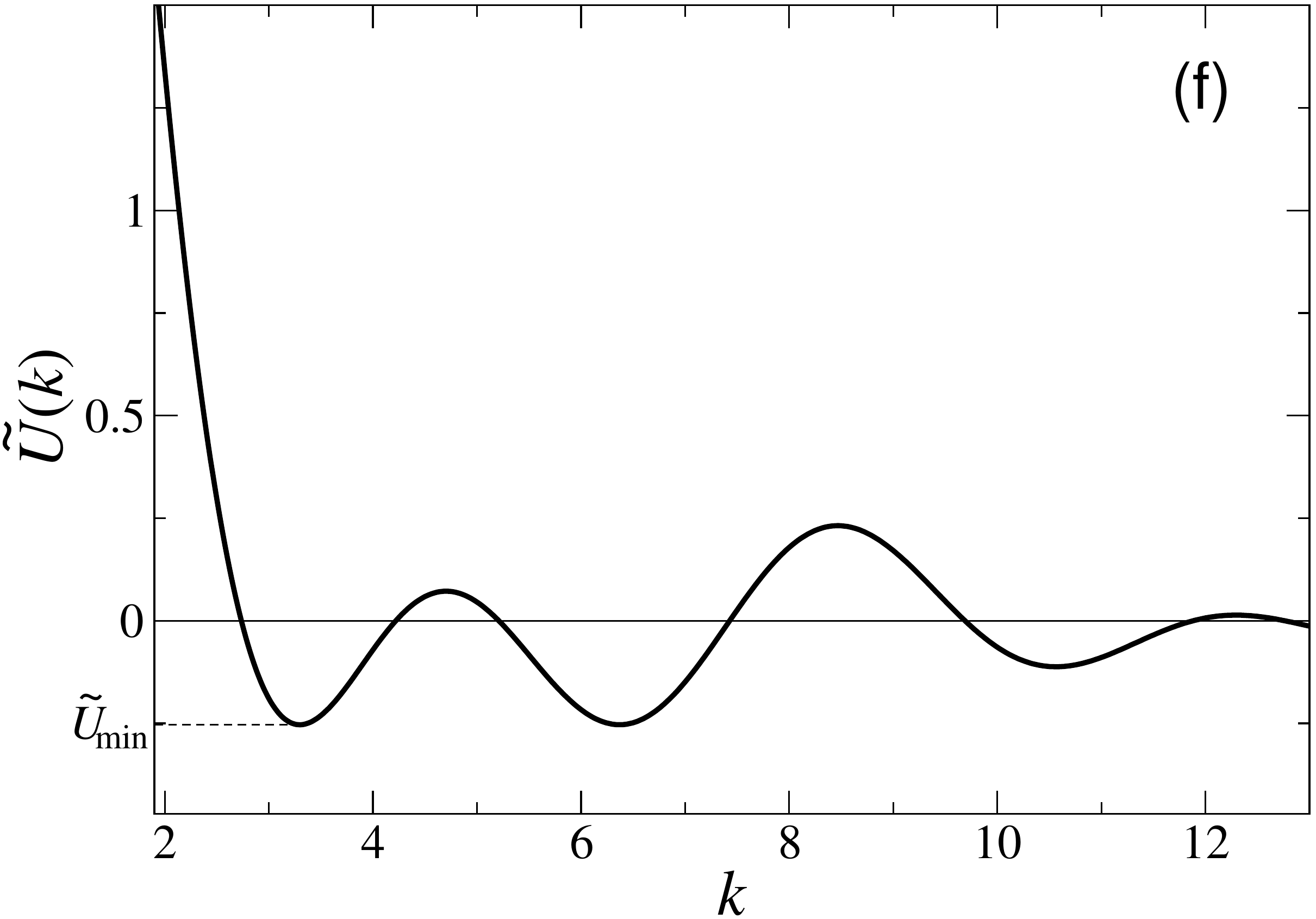}}}\\
\end{center}
\caption{Three of the isotropic pair-potentials used in this study
  (top row), whose Fourier transforms (bottom row) are designed so
  that the ratio $q$ of the positions of their first two minima is
  approximately 1.93. The first potential in (a), motivated by the
  structure of dendritic micelles, consists of a normalized repulsive
  core, with a short range (van der Waals) attraction at the core
  boundary, and an exponentially-decaying soft shoulder. This behavior
  is approximated by even simpler 3-step and 2-step potentials, shown
  in (c) and (e). \emph{Potential parameters}: (a) $U(r)=1$ for
  $0<r<1$; and $-u_1/(r-1+\delta) + u_2\exp[-(r-1)/\lambda]$, with
  $u_1=0.02057$, $u_2=1.057$, $\lambda=0.9756$, and $\delta=0.01$, for
  $r>1$. (c) $U(r)=1$ for $0<r<1$; $-0.0770$ for $1<r<1.0485$;
  $0.4690$ for $1.0485<r<1.6570$; and 0 for $r>1.6570$. (e) $U(r)=1$
  for $0<r<1$; $0.457$ for $1<r<1.7442$; and 0 for $r>1.7442$. }
  \label{fig:potentials}
\end{figure*}
%##################### Figure 1 #################################### 

To try to predict the conditions under which dodecagonal quasicrystals
minimize $\cF$ of Eq.~(\ref{cF}), let us recall the findings of LP,
who showed that such crystals are stabilized in their simpler model,
$\cFLP$ of Eq.~(\ref{eq:FLP}), under the following conditions: (i)
$q\simeq 2\cos(\pi/12)$; (ii) $0<\varepsilon/\alpha^2\lesssim 0.088$.
LP further showed that, for $\varepsilon/\alpha^2\gtrsim 0.088$, a
hexagonal state is obtained. In terms of our free energy
expansion~(\ref{eq:Fexpansion}), condition (i) translates into a
requirement for the properties of $\tU(\veck)$. The first two minima
of this function should be located at wavevectors $k_1$ and $k_2$,
whose ratio $q=k_2/k_1$ is around $2\cos(\pi/12)\simeq 1.93$, and
should have a similar depth.  Condition (ii) translates into a
requirement for the thermodynamic parameters, $T$ and $\mu$ (or
$\cbar$), namely, $0<1-T/\Tc\lesssim 0.066$, {\it i.e.\/} $T$ is
restricted to a small range below $\Tc$.

We test these estimates, obtained from the approximate free energy
expansion~(\ref{eq:Fexpansion}), against the numerical minimization of
$\cF\left[c(\vecr)\right]$ of Eq.~(\ref{cF}).  We find that a variety
of isotropic pair potentials, three examples of which are shown in
Fig.~\ref{fig:potentials}, contain a sufficient number of tunable
parameters to satisfy condition (i) above. The form of these
potentials is motivated by the qualitative features of the
experimental system,\cite{zeng04,*zeng05} consisting of spherical
micelles whose interaction should contain an inner repulsive core, a
region governed by van der Waals attraction, and a longer-range soft
repulsion. Using these potentials in Eq.~(\ref{cF}), we find the
minimum free-energy state by numerically integrating the corresponding
relaxational equation, $\partial_t c = -\delta\cF / \delta c$, using a
pseudospectral method---with local terms evaluated in direct space
and non-local terms in Fourier space---starting with random initial
conditions and waiting until a steady state is obtained. We note that
this equation does not describe the actual dynamics and is used merely
as a minimization tool.

\begin{figure*}[t]
\begin{center}
\scalebox{0.235}{\rotatebox{00}{\includegraphics{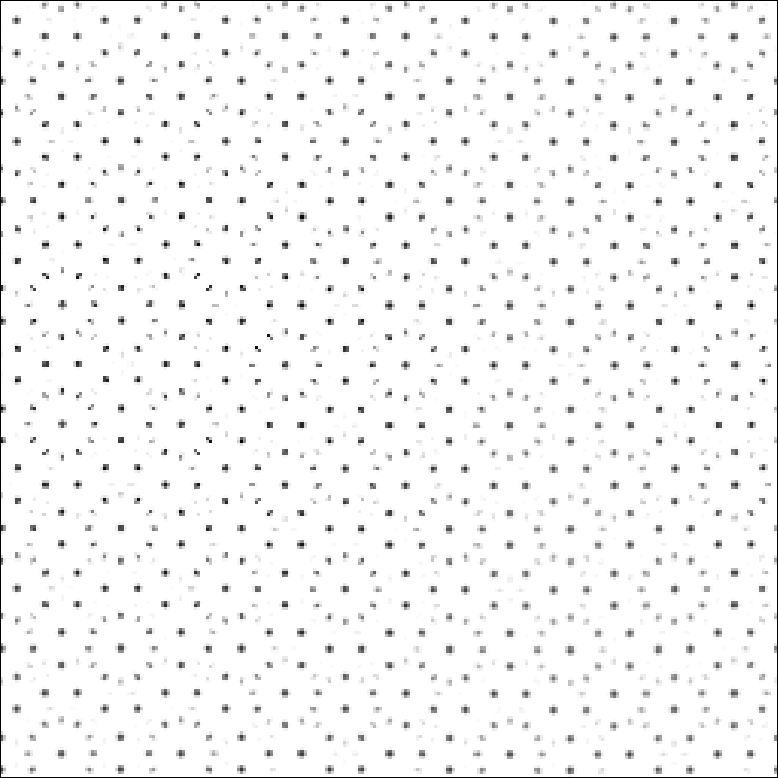}}}
\scalebox{0.235}{\rotatebox{00}{\includegraphics{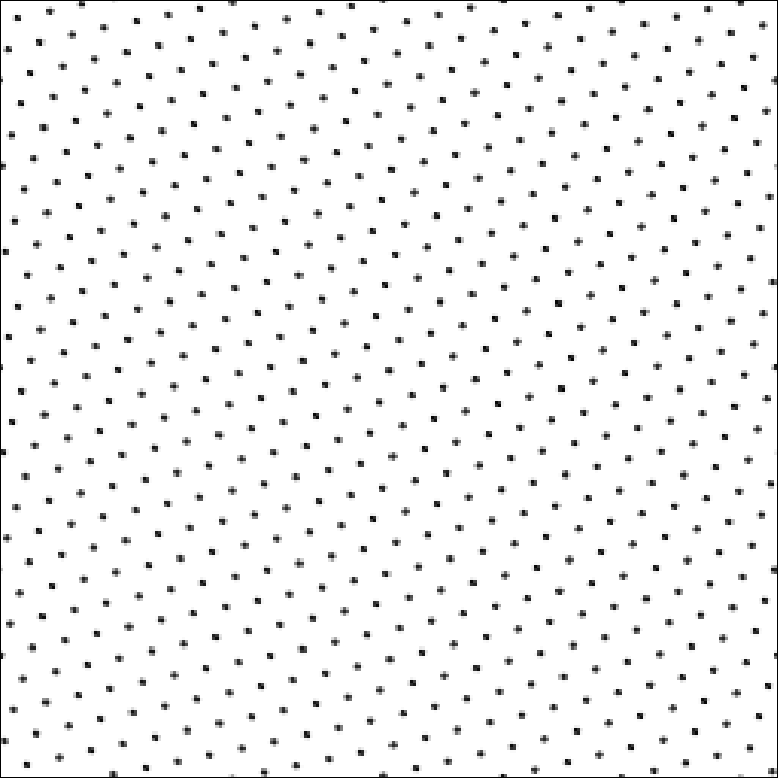}}}
\scalebox{0.235}{\rotatebox{00}{\includegraphics{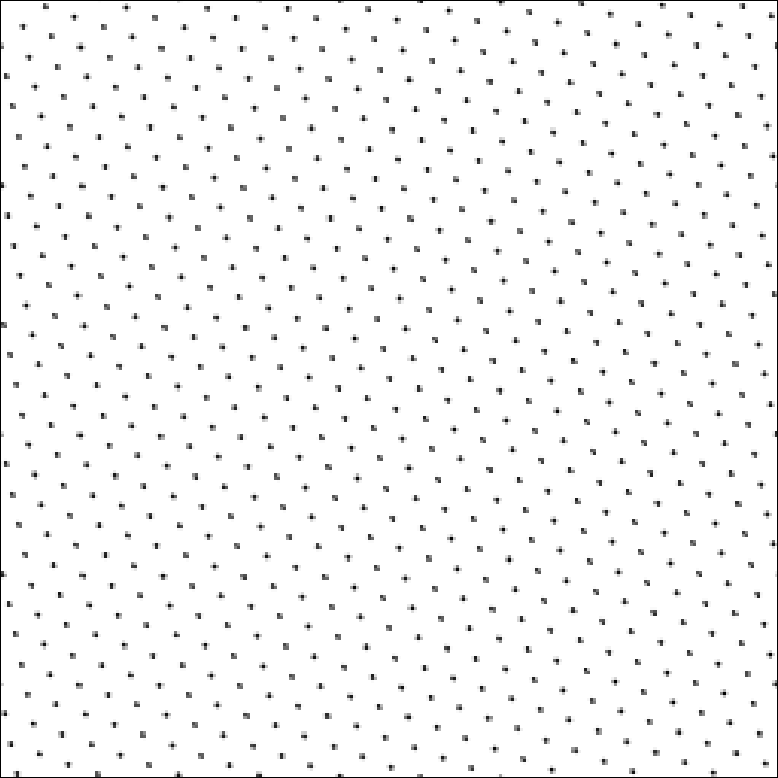}}}\\\vskip1pt
\scalebox{0.235}{\rotatebox{00}{\includegraphics{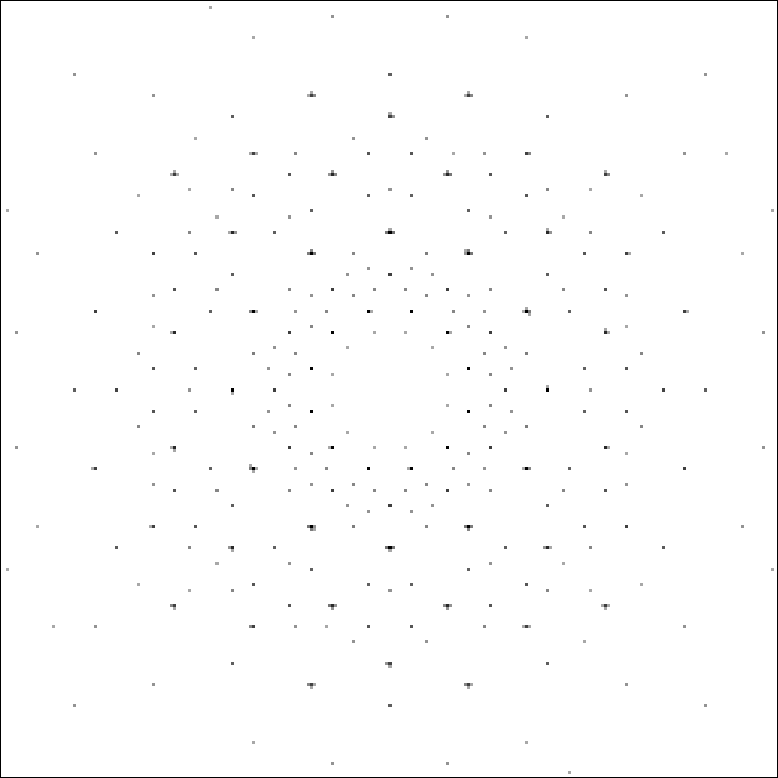}}}
\scalebox{0.235}{\rotatebox{00}{\includegraphics{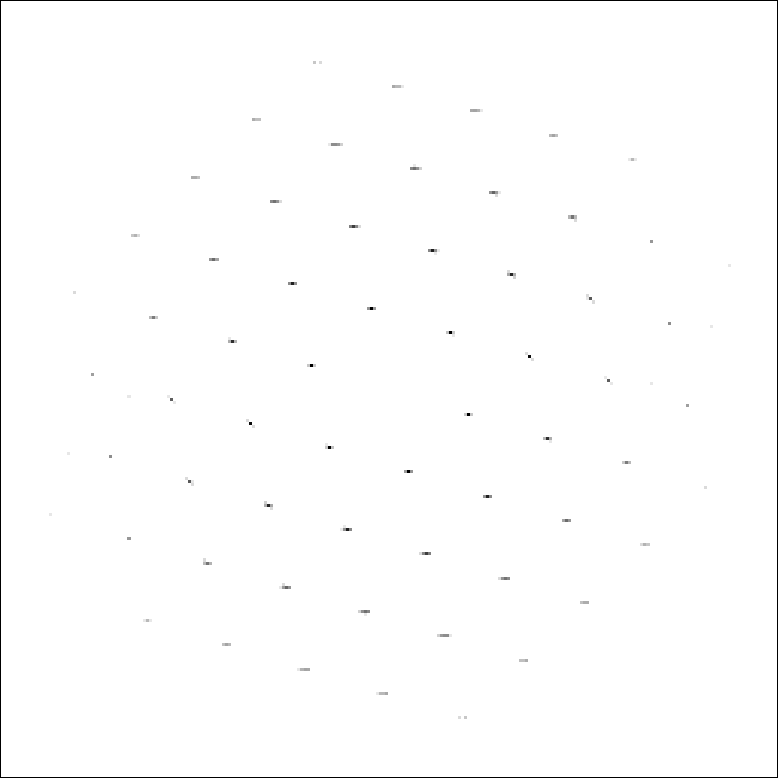}}}
\scalebox{0.235}{\rotatebox{00}{\includegraphics{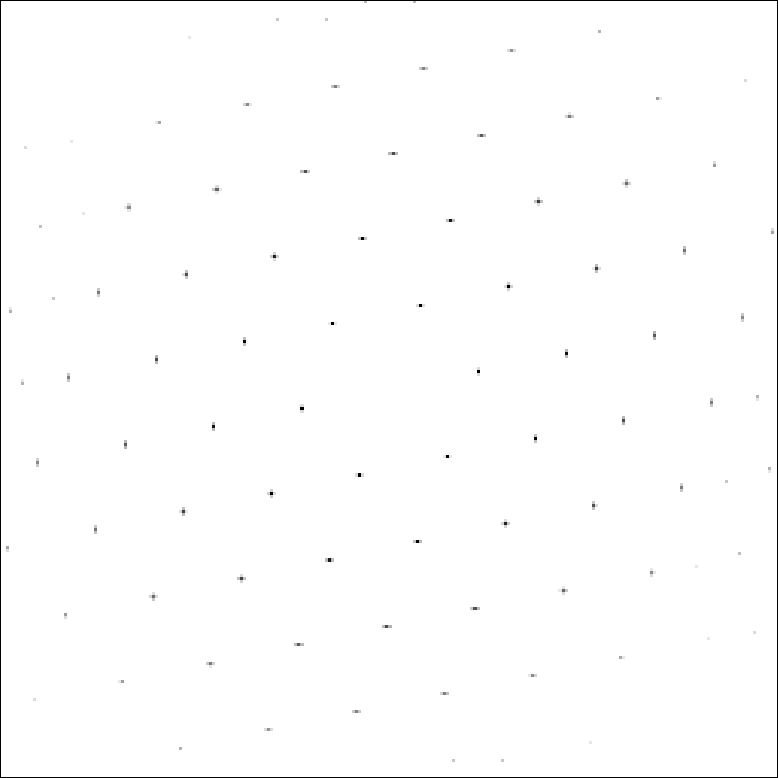}}}\\
\end{center}
\caption{Real space densities $c(\vecr)$ (top row) and their Fourier
  transforms $|c(\veck)|$, with $c(\veck=0)$ removed (bottom row),
  obtained by a numerical minimization of the free energy~(\ref{cF}),
  with the 3-step potential of Fig.~\ref{fig:potentials}. Other
  potentials yield similar results. \emph{Dodecagonal crystal} (left
  column): 3-step potential, with parameters specified in
  Fig.~\ref{fig:potentials}(c). Thermodynamic variables are
  $\mu=2.2895$ and $T=0.999\Tc$. \emph{Hexagonal crystal} (middle
  column): Same as in the left column, but with a lower temperature
  $T=0.980\Tc$.  \emph{Hexagonal crystal} (right column): 3-step
  potential similar to Fig.~\ref{fig:potentials}(c), with parameters
  $U(r)=1$ for $0<r<1$; $-0.115$ for $1<r<1.053$; $0.370$ for
  $1.053<r<1.600$; and 0 for $r>1.600$, designed so that
  $q\simeq\sqrt3$.  Thermodynamic variables are $\mu=1.9041$ and
  $T=0.999\Tc$.}
  \label{fig:results}
\end{figure*}

When $q$ is selected according to condition (i)---as demonstrated in
Fig.~\ref{fig:potentials}---and $T$ and $\mu$ are set such that $T$ is
just below $\Tc$ [condition (ii)], we indeed find a minimum
free-energy state which is a dodecagonal quasicrystal, as shown on the
left-hand column of Fig.~\ref{fig:results} for the particular case of
the 3-step potential of Fig.~\ref{fig:potentials}(c).  When $T$ is
decreased slightly further a transition is observed to a hexagonal
state [middle column of Fig.~\ref{fig:results}], restricting the
stability of dodecagonal quasicrystals to a narrower range than that
predicted by the approximate free energy
expansion~(\ref{eq:Fexpansion}).\footnote{This narrower stability
  region, however, may be a consequence of our mean-field
  coarse-graining and should be checked in a more detailed future
  study.}  The same stabilizing mechanism of two length scales and
3-body interactions can be used to obtain other
structures;\cite{faraday} for example, setting $q=\sqrt{3}$ yields
immediately below $\Tc$ a hexagonal crystal, shown on the right-hand
column of Fig.~\ref{fig:results}. We have observed that slight
variations in the potential parameters still yield the expected
phases. A systematic study of the stability boundaries in parameter
space is left for future work, perhaps using more realistic
potentials, derived from specific experimental realizations.

%##################### Figure 2 #################################### 
%##################### Figure 2 #################################### 

While the first two minima of the Fourier-transformed pair potential
$\tU(k)$ must be negative to give a positive $\Tc$, the value of
$\tU(k=0)$ need not be negative. This allows a purely repulsive pair
potential $U(r)$ to stabilize a quasicrystal, as we show here for the
simple two-step potential of Fig.~\ref{fig:potentials}(e). Despite the
appealing simplicity of this potential---defined after scaling by two
parameters only---we emphasize that potentials of this sort can
satisfy the requirements on the Fourier space minima only near a
single choice of their two parameters. As a consequence, attempting a
na\"{\i}ve numerical search for a dodecagonal quasicrystal with this
potential, by setting the extent of the second step to
$R=2\cos(\pi/12)\simeq1.93$, would fail. We now understand that in
order for nonlinear mode interactions to stabilize a dodecagonal
quasicrystal, one must adjust this ratio not in the real-space pair
potential, but rather in its Fourier transform. This is obtained in
this case by setting $R\simeq1.74$ in real space. We have managed
to stabilize a dodecagonal quasicrystal with the simple two-step
potential by eliminating the need to search around in parameter
space. Our theoretical understanding of the source of stability allows
us to point at the stability regions in parameter space, even when
these regions are extremely narrow.

Thus, we confirm that the existence of two characteristic length
scales and sufficiently strong 3-body or higher-order nonlinear
interactions can account for the stability of dodecagonal
quasicrystals of isotropic soft particles.  More specifically, we
determine the two length scales through requirements on the minima of
the Fourier-transformed pair potential and, although 3-body terms may
arise from various interactions,\cite{vonFerber00} we show that
translational entropy suffices to provide the required term. Thus, the
delicate interplay between interaction and entropy can give rise to
stable quasicrystals even for relatively simple isotropic potentials.
By designing these potentials along the guidelines provided here, one
should be able to control the self-assembly of quite complex
structures. Preliminary tests, using molecular dynamics
simulations, indicate that our design principles seem to
work.\cite{EngelPrivate} 

This work can be extended in several directions. The results of the
approximate theory presented here should be verified using direct
computer simulations. Importantly, such simulations can be used to
clarify the actual stability regions of the quasicrystalline phase.
The dynamics of crystallization, and of collective degrees of freedom
in the ordered state, can be studied by replacing the equation used
here to minimize the free energy by one that is adequate for a
conserved density field, possibly while taking thermal fluctuations
into account. Finally, a similar coarse-graining procedure could be
applied to two-component systems or anisotropic
potentials.\cite{hayashida07,talapin09}

We are grateful to Michael Cross, Michael Engel, and Moshe Schwartz
for fruitful discussions. This research is supported by the Israel
Science Foundation through Grants No.~684/06 and 556/10.

\bibliography{soft}

\end{document}